\documentclass[twocolumn]{jpsj2}
\usepackage{graphicx}
\usepackage{bm}

\title{Duality Relations and Exotic Orders in Electronic
Ladder Systems}

\author{Tsutomu Momoi and
Toshiya Hikihara\thanks{Present address: Division of Physics,
Graduate School of Science, Hokkaido University, Sapporo 060-0810,
Japan}}

\inst{Condensed Matter Theory Laboratory, RIKEN, Wako, Saitama
351-0198, Japan}

\abst{We discuss duality relations in correlated electronic ladder
systems to clarify mutual relations between various conventional
and unconventional phases.
For the generalized two-leg Hubbard ladder, we find two exact
duality relations, and also one asymptotic relation which holds in
the low-energy regime. These duality relations show that
unconventional (exotic) density-wave orders such as staggered flux
or circulating spin-current are directly mapped to conventional
density-wave orders, which establishes the appearance of various
exotic states with time-reversal and/or spin symmetry breaking. We
also study duality relations in the SO(5) symmetry that was
proposed to unify antiferromagnetism and $d$-wave
superconductivity. We show that the same SO(5) symmetry also
unifies circulating spin current order and $s$-wave
superconductivity.}

\kword{Electronic ladder, Hubbard model, duality relation, exotic
order, SO(5) symmetry, staggered flux, spin circulating current.}

\begin{document}

\maketitle

\section{Introduction}
Unconventional density-wave orders such as $d$-density wave
($d$DW) (which is equivalently called as staggered flux or orbital
antiferromagnets) and $d$-spin-density wave ($d$SDW) (which is
circulating spin current) were first proposed in the context of
excitonic insulators\cite{HalperinR} and later discussed in
high-$T_c$ superconductors\cite{AffleckM,Schultz,NersesyanV}, but
the appearance of these orders was not established at that time.
Recently, several experimental results have led to a resurgence of
interest in the possibility of these exotic orders. A $d$DW
state\cite{ChakravartyLMN} was discussed to appear in the
underdoped region of high-$T_c$ superconductors, where a pseudogap
was observed\cite{TimuskS}, and also in the low-temperature phase
of the quasi-two-dimensional organic conductor\cite{DoraMV},
$\alpha$-(BEDT-TTF)$_2$KHg(SCN)$_4$. Both a $d$SDW
state\cite{IkedaO} and a $d$DW state\cite{ChandraCMT} were
proposed as origins of hidden order in the heavy-fermion compounds
URu$_2$Si$_2$ and UPt$_3$.

The appearance of unconventional (exotic) orders has been tested
in microscopic models of correlated electrons. In particular, the
generalized Hubbard model on the two-leg ladder has been
attracting attention as a minimal model for showing the exotic
orders\cite{LinBF,MarstonFS,FjaerestadM,TsuchiizuF,WuLiuF,Schollwoeck}.
As is well known, the standard two-leg Hubbard and $t$-$J$ ladders
without any inter-site interaction do not show any
order\cite{DagottoR,MarstonFS,Schollwoeck}. Both strong-coupling
and weak-coupling analyses however found that the generalized
Hubbard ladder model with inter-site interactions exhibits various
phases at half-filling\cite{LinBF,FjaerestadM,TsuchiizuF,WuLiuF}.
There are at least eight phases\cite{TsuchiizuF}, i.e.,
charge-density-wave (CDW), $d$DW, $p$-density wave ($p$DW),
$f$-density wave ($f$DW), $d^{(')}$-Mott, and $s^{(')}$-Mott
phases. For less than half-filling, large-scale numerical
calculations also reported CDW\cite{VojtaHN} and
$d$DW\cite{Schollwoeck} phases. Bosonaization and
renormalization-group analysis revealed the appearance of
quasi-long-range order of density waves\cite{WuLiuF,OrignacC}.
Recently we found two exact duality relations between these
various phases\cite{MomoiH}. The relations show that
unconventional density-wave orders such as staggered flux or
circulating spin current are dual to conventional density-wave
orders and there are direct one-to-one mappings between dual
phases in the generalized Hubbard ladder systems.

The electronic ladder system also serves as a playground for
theories of high-$T_c$ superconductivity. The SO(5) theory, which
was proposed to unify antiferromagnetism (AFM) and $d$-wave
superconductivity ($d$SC) in terms of the SO(5)
symmetry\cite{SCZhang,RabelloKDZ,Henley}, was also applied to the
Hubbard ladder. Scalapino {\it et al.}\cite{ScalapinoZH} and later
many groups\cite{MarstonFS,LinBF,FrahmS,FjaerestadM} studied the
Hubbard ladder with the SO(5) symmetry to capture some of the
basic low-energy physics of the high-$T_c$ cuprates. There, one
grand order parameter field
\begin{equation}
(\sqrt{2}{\rm Re}{{\cal O}_{d{\rm SC}}},{\cal N}_x,{\cal
N}_y,{\cal N}_z,\sqrt{2}{\rm Im}{{\cal O}_{d{\rm SC}}})
\label{eq:SO5vec1}
\end{equation}
behaves as a five-component vector, where ${\cal O}_{d{\rm SC}}$
denotes the pairing operator of $d$SC with ${\rm Re}{\cal
O}=\frac{1}{2}({\cal O}^\dagger+{\cal O})$, ${\rm Im}{\cal
O}=\frac{1}{2i}({\cal O}^\dagger-{\cal O})$, and the three
elements ${\cal N}_\alpha$ with $\alpha=x,y,z$ are Cartesian
components of the staggered magnetization.

\begin{figure}[tb]
  \centering
\includegraphics[width=83mm]{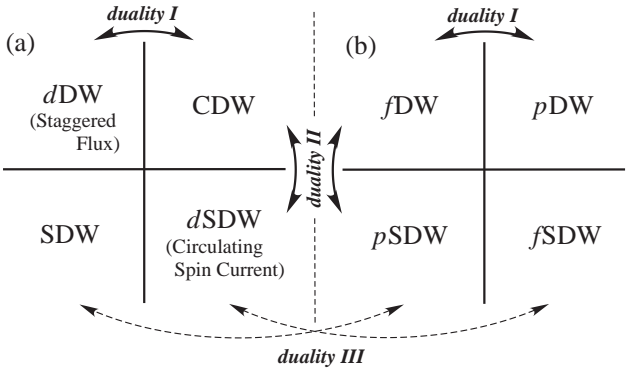}
\caption{Two exact duality relations (I and II) between four
phases with (a) CDW, SDW, $d$DW, and $d$SDW orders, and (b) $p$DW,
$f$DW, $p$SDW, and $f$SDW orders. There is also another duality
relation (III), which appears only in the low-energy region at
half-filling.}\label{fig:duality}
\end{figure}

In this paper we develop further the duality relations given in
ref.\ \citen{MomoiH} and also study the duality structure in the
SO(5) symmetric Hubbard ladder. First, we discuss two exact
duality relations in the generalized Hubbard ladder. One duality
transformation relates conventional density waves to
unconventional density (or current) waves and the other relates
charge-density degrees of freedom to spin-current ones. These
transformations also show duality relations between $s$- and
$d$-wave superconductivity ($s$SC and $d$SC).
Furthermore, we show another asymptotic duality relation, which
appears only in the low-energy effective theory at half-filling.
These duality relations among various density-wave phases are
summarized in Fig.\ \ref{fig:duality}. The transformations give
one-to-one parameter mappings between dual phases. If one finds an
ordered phase in a certain parameter space, one can naturally
conclude the appearance of dual ordered phases in dual parameter
spaces. These relations help us to find various new exotic phases
with spin and/or time-reversal symmetry breaking.

Next, we apply the duality relation to the SO(5) symmetry. We find
that the SO(5) symmetry also unifies $d$SDW (circulating spin
current) order and $s$-wave SC. Also, the SO(5) symmetric Hubbard
ladder has a duality structure. In one parameter region, the
original AFM-$d$SC SO(5) vectors represent dominant correlations,
but in the other dual region the new $d$SDW and $s$SC vectors are
dominant showing a crossover from $d$SDW to $s$SC upon doping.

This paper is structured as follows. The definitions of the
Hamiltonian and operators are given in Sec.\ 2. The two exact
duality relations I and II are described in Sec.\ 3. The duality
structure in the generalized Hubbard ladder model is shown in
Sec.\ 4. Section 5 contains a discussion on the duality in the
SO(5) symmetry. Duality relations are reconsidered by using the
Bosonization framework in Sec.~6. An application of the duality
relations to results in the strong-coupling limit is discussed in
Sec.\ 7. Section 8 contains discussions.

\section{Definitions}
\subsection{Hamiltonian}
We consider the generalized two-leg Hubbard ladder, which contains
the on-site repulsion $U$, the intra-rung repulsion $V_\perp$, the
intra-rung spin-exchange $J_\perp$ ($J_\perp^z$), and the
intra-rung pair hopping $t_{\rm pair}$. The Hamiltonian is given
by
\begin{equation}
H = H_{0} + H_{\rm rung}-\mu\sum_{j,l,\sigma} n_{j,l,\sigma}
\label{eq:Ham}
\end{equation}
\begin{subequations}
with
\begin{align}
H_{0} &= - \sum_{j,\sigma}[\{ t_\parallel (c_{j,1,\sigma}^\dagger
c_{j+1,1,\sigma} + c_{j,2,\sigma}^\dagger
c_{j+1,2,\sigma})\nonumber\\
&+ t_\perp c_{j,1,\sigma}^\dagger c_{j,2,\sigma}\} + {\rm H.c.}],
\label{eq:Ht} \\
H_{\rm rung} &= \sum_{j} \Biggl[ U \sum_{l} n_{j,l,\uparrow}
n_{j,l,\downarrow} + V_\perp \sum_{\sigma,\sigma'} n_{j,1,\sigma}
n_{j,2,\sigma'} \nonumber\\
&+ J_\perp (S^x_{j,1} S^x_{j,2}+S^y_{j,1}
S^y_{j,2})+ J^z_\perp S^z_{j,1} S^z_{j,2}\nonumber \\
&+ t_{\rm pair} (c_{j,1,\uparrow}^\dagger
c_{j,1,\downarrow}^\dagger c_{j,2,\downarrow} c_{j,2,\uparrow} +
{\rm H.c.}) \Biggr]. \label{eq:Hrung}
\end{align}
\end{subequations}
Here $c^\dagger_{j,l,\sigma}$ ($c_{j,l,\sigma}$) ($l=1,2$ and
$\sigma=\uparrow,\downarrow$) denotes an electron creation
(annihilation) operator on the $l$th site of the $j$th rung,
$n_{j,l,\sigma} = c_{j,l,\sigma}^\dagger c_{j,l,\sigma}$ and ${\bm
S}_{j,l} = \frac{1}{2} \sum_{\sigma,\sigma'}
c_{j,l,\sigma}^\dagger {\bm \sigma}_{\sigma\sigma'}
c_{j,l,\sigma'}$ with the Pauli matrices $\sigma^\alpha$
($\alpha=x,y,z$). In this paper we consider both spin isotropic
($J_\perp=J^z_\perp$) and anisotropic cases.

\subsection{Order operators and typical states}
Here, we list the definition of order operators used in the
present work.
We consider a charge-density-wave (CDW) operator (or
$s$-density-wave operator)
\begin{equation}
{\cal O}_{\rm CDW}(j) = \frac{1}{2} \sum_\sigma
(n_{j,1,\sigma}-n_{j,2,\sigma}),\label{eq:O_PDW}
\end{equation}
a $d$-density-wave ($d$DW) operator
\begin{equation}
{\cal O}_{d{\rm DW}}(j) = \frac{i}{2} \sum_\sigma
(c_{j,1,\sigma}^\dagger c_{j,2,\sigma} - {\rm H.c.}),
\end{equation}
a $p$-density-wave ($p$DW) operator (or staggered dimer operator)
\begin{equation}
{\cal O}_{p{\rm DW}}(j) = \frac{1}{4} \sum_\sigma
[(c_{j+1,1,\sigma}^\dagger c_{j,1,\sigma} -
c_{j+1,2,\sigma}^\dagger c_{j,2,\sigma})+ {\rm H.c.} ],
\end{equation}
and an $f$-density-wave ($f$DW) operator (or diagonal current
operator)
\begin{equation}
{\cal O}_{f{\rm DW}}(j) = \frac{i}{4} \sum_\sigma
[(c_{j+1,1,\sigma}^\dagger c_{j,2,\sigma} -
c_{j+1,2,\sigma}^\dagger c_{j,1,\sigma}) -{\rm H.c.} ].
\label{eq:O_FDW}
\end{equation}
In the similar way, we consider operators in the spin sector.
Inserting the Pauli matrix $\sigma^z_{\sigma\sigma}$ into the
right-hand sides of eqs.\ (\ref{eq:O_PDW})-(\ref{eq:O_FDW}), one
defines a spin-density-wave (SDW) operator
\begin{equation}
{\cal O}_{\rm SDW}(j) = \frac{1}{2} \sum_\sigma \sigma^z_{\sigma
\sigma} (n_{j,1,\sigma}-n_{j,2,\sigma}),
\end{equation}
a $d$-spin-density-wave ($d$SDW) operator
\begin{equation}
{\cal O}_{d{\rm SDW}}(j) = \frac{i}{2} \sum_\sigma
\sigma^z_{\sigma \sigma} (c_{j,1,\sigma}^\dagger c_{j,2,\sigma} -
{\rm H.c.}),
\end{equation}
a $p$-spin-density wave ($p$SDW) operator ${\cal O}_{p{\rm
SDW}}(j)=\frac{1}{4} \sum_\sigma \sigma^z_{\sigma
\sigma}[(c_{j+1,1,\sigma}^\dagger c_{j,1,\sigma} -
c_{j+1,2,\sigma}^\dagger c_{j,2,\sigma})+ {\rm H.c.}]$, and an
$f$-spin-density wave ($f$SDW) operator ${\cal O}_{f{\rm
SDW}}(j)=\frac{1}{4} \sum_\sigma \sigma^z_{\sigma
\sigma}[(c_{j+1,1,\sigma}^\dagger c_{j,2,\sigma} -
c_{j+1,2,\sigma}^\dagger c_{j,1,\sigma})+ {\rm H.c.} ]$, 
respectively. Note that the CDW and $p$DW (SDW and $p$SDW) orders
are kinds of density waves of charges (spins) while the $d$DW and
$f$DW ($d$SDW and $f$SDW) orders finite local charge (spin)
currents. Accurately speaking, $d$DW and $f$DW operators do not
correspond to exact current operators if there are pair-hopping
terms ($t_{\rm pair}\ne 0$) in the Hamiltonian, but they can
detect time-reversal symmetry breaking, which is closely related
to currents. Order parameters on the ladder are given by
\[
O_A(q)=L^{-1} \sum_j {\cal O}_{A}(j)\exp(iqj),
\]
where $L$ is the number of rungs.

Moreover, we consider the $d$- and $s$-wave pairing operators,
\begin{align}
{\cal O}_{d{\rm SC}} (j) &= \frac{1}{\sqrt{2}}
(c_{j,1,\uparrow}c_{j,2,\downarrow} - c_{j,1,\downarrow}
c_{j,2,\uparrow}),
\\
{\cal O}_{s{\rm SC}} (j) &= \frac{1}{\sqrt{2}}
(c_{j,1,\uparrow}c_{j,1,\downarrow} + c_{j,2,\uparrow}
c_{j,2,\downarrow}),
\end{align}
which characterize the $d$- and $s$-wave spin-singlet
superconductivity ($d$SC and $s$SC), respectively. Using these
operators, the representatives of the $d$- and $s$-wave
Mott-insulating ($d$- and $s$-Mott) states at half-filling are
given by
\begin{align}
|\mbox{$d$-Mott} \rangle = \prod_j {\cal O}^\dagger_{d{\rm SC}}
(j) |0\rangle,~~~~~ |\mbox{$s$-Mott} \rangle = \prod_j {\cal
O}^\dagger_{s{\rm SC}} (j) |0\rangle,\nonumber
\end{align}
respectively, where $|0 \rangle$ is the vacuum of the electron
operators. We also consider the spin-triplet and spin-singlet
pairing operators with odd parity given by
\begin{align}
{\cal O}_{\rm t,o}(j) &= \frac{1}{\sqrt{2}}(c_{j,1,\uparrow}
c_{j,2,\downarrow} + c_{j,1,\downarrow} c_{j,2,\uparrow}),
\label{eq:o_to_pairing}\\
{\cal O}_{\rm s,o}(j) &= \frac{1}{\sqrt{2}}(c_{j,1,\uparrow}
c_{j,1,\downarrow} - c_{j,2,\uparrow} c_{j,2,\downarrow}).
\label{eq:o_so_pairing}
\end{align}


\section{Exact duality transformations I and II}
Here we describe the duality transformations on electron
operators\cite{MomoiH} so that this paper is self-contained. They
are given by gauge transformations on the bonding and antibonding
operators $d_{j,\pm,\sigma} = (c_{j,1,\sigma} \pm
c_{j,2,\sigma})/\sqrt{2}$. Two duality transformations are
presented in the following.

\subsection{Duality relation I: density and current}

Consider a gauge transformation of antibonding operators given by
the unitary operator
\begin{equation}
U_I(\theta)=\prod_{j,\sigma} \exp ( -i \theta
d^\dagger_{j,-,\sigma} d_{j,-,\sigma} ).\label{eq:unitaryI}
\end{equation}
In the case of $\theta=\pi/2$, this operator gives the duality
transformation I, $\tilde{d}_{j,\pm,\sigma} \equiv U_I(\pi/2)d_{
j,\pm,\sigma}{U_I}(\pi/2)^{-1}$, which yields
\begin{equation}
\tilde{d}_{j,+,\sigma} = d_{j,+,\sigma},~~~~~~~~
\tilde{d}_{j,-,\sigma} = i d_{j,-,\sigma} \label{eq:dual1}
\end{equation}
for $\sigma=\uparrow,\downarrow$. In terms of the electron
operators $c_{j,l,\sigma}$, this transformation is written as
\begin{equation}
\begin{split}
\tilde{c}_{j,1,\sigma} &= (e^{\pi i/4} c_{j,1,\sigma} + e^{-\pi
i/4} c_{j,2,\sigma})/\sqrt{2},\\
\tilde{c}_{j,2,\sigma} &= (e^{-\pi i/4} c_{j,1,\sigma} + e^{\pi
i/4} c_{j,2,\sigma})/\sqrt{2}. \label{eq:tra1}
\end{split}
\end{equation}


Applying the transformation (\ref{eq:dual1}), we can obtain dual
representations of operators straightforwardly as follows: The
density-wave operators are transformed as
\begin{align}
\tilde{\cal O}_{\rm CDW} &= - {\cal O}_{d{\rm DW}},~~~~~~~
\tilde{\cal O}_{d{\rm DW}}={\cal O}_{\rm CDW},
\nonumber\\
\tilde{\cal O}_{\rm SDW} &= - {\cal O}_{d{\rm SDW}},~~~~
\tilde{\cal O}_{d{\rm SDW}}={\cal O}_{\rm SDW},\nonumber\\
\tilde{\cal O}_{p{\rm DW}} &= - {\cal O}_{f{\rm DW}},~~~~~~~
\tilde{\cal O}_{f{\rm DW}} = {\cal O}_{p{\rm DW}}, \nonumber\\
\tilde{\cal O}_{p{\rm SDW}} &= - {\cal O}_{f{\rm SDW}},~~~~
\tilde{\cal O}_{f{\rm SDW}} = {\cal O}_{p{\rm SDW}}.
\label{eq:dualityI}
\end{align}
Thus, current-order operators, such as $d$DW, $d$SDW, $f$DW, and
$f$SDW operators, turn out to be dual to density-wave operators,
CDW, SDW, $p$DW, and $p$SDW operators, respectively.
It should be noted that the unitary operator $U_I(\theta)$ gives a
continuous transformation between dual operators. For example, if
one considers CDW and $d$DW operators, they are transformed as
\begin{equation}
U_I (\theta){\cal O}_{\rm CDW}U_I (\theta)^{-1}= {\cal O}_{\rm
CDW}\cos \theta - {\cal O}_{d{\rm DW}}\sin \theta.
\label{eq:dual1_cont}
\end{equation}

The $s$- and $d$-wave pairing operators are also related by the
transformation (\ref{eq:dual1}) as
\begin{equation}
\tilde{\cal O}_{s{\rm SC}} = {\cal O}_{d{\rm SC}}.
\end{equation}
Thus the $d$-wave SC phase is dual to the $s$-wave SC phase. By
definitions, $d$-Mott and $s$-Mott phases are also dual to each
other,
\begin{equation}
|\widetilde{\mbox{$d$-Mott}}\rangle = |\mbox{$s$-Mott}\rangle,
\end{equation}
where we omitted a constant factor. On the other hand, the
parity-odd pairing operators (\ref{eq:o_to_pairing}) and
(\ref{eq:o_so_pairing}) are invariant under the transformation.

\subsection{Duality relation II: density and spin current}

Consider a gauge transformation given by the unitary operator
\begin{equation}\label{eq:unitaryII}
U_{II}(\theta)=\prod_j \exp \{ -i \theta (d^\dagger_{j,+,\uparrow}
d_{j,+,\uparrow} + d^\dagger_{j,-,\downarrow}
d_{j,-,\downarrow})\}.
\end{equation}
This unitary with $\theta=\pi/2$ gives the duality transformation
II, $\bar{d}_{j,\pm,\sigma} \equiv U_{II}(\pi/2)d_{
j,\pm,\sigma}{U_{II}}(\pi/2)^{-1}$, which yields
\begin{equation}
\begin{array}{lc}
  \bar{d}_{j,+,\uparrow} = i d_{j,+,\uparrow},~~~~~~
  & \bar{d}_{j,+,\downarrow} = d_{j,+,\downarrow}, \\
  \bar{d}_{j,-,\uparrow} = d_{j,-,\uparrow},
  & \bar{d}_{j,-,\downarrow} = i d_{j,-,\downarrow}.\\
\end{array}
\label{eq:dual2}
\end{equation}
In terms of the electron operators $c_{j,l,\sigma}$, this
transformation is written as
\begin{align}
\bar{c}_{j,1,\uparrow} &= (e^{\pi i/4} c_{j,1,\uparrow} +
e^{3\pi i/4} c_{j,2,\uparrow})/\sqrt{2},
\nonumber \\
\bar{c}_{j,2,\uparrow} &= (e^{3 \pi i/4} c_{j,1,\uparrow} +
e^{\pi i/4} c_{j,2,\uparrow})/\sqrt{2},
\nonumber \\
\bar{c}_{j,1,\downarrow} &= (e^{\pi i/4} c_{j,1,\downarrow} +
e^{-\pi i/4} c_{j,2,\downarrow})/\sqrt{2}, \nonumber\\
\bar{c}_{j,2,\downarrow} &= (e^{-\pi i/4} c_{j,1,\downarrow} +
e^{\pi i/4} c_{j,2,\downarrow})/\sqrt{2}. \label{eq:tra2}
\end{align}

It is easily shown that the transformation (\ref{eq:dual2}) gives
duality relations between density-waves operators as follows
\begin{align}
\bar{\cal O}_{\rm CDW} &= {\cal O}_{d{\rm SDW}},~~~~~~~
\bar{\cal O}_{d{\rm SDW}}=-{\cal O}_{\rm CDW},\nonumber\\
\bar{\cal O}_{d{\rm DW}} &= -{\cal O}_{\rm SDW},~~~~~~~
\bar{\cal O}_{\rm SDW}={\cal O}_{d{\rm DW}},\nonumber\\
\bar{\cal O}_{p{\rm DW}} &= {\cal O}_{f{\rm SDW}},~~~~~~~
\bar{\cal O}_{f{\rm SDW}}=-{\cal O}_{p{\rm DW}},\nonumber\\
\bar{\cal O}_{f{\rm DW}} &= -{\cal O}_{p{\rm SDW}},~~~~~
\bar{\cal O}_{p{\rm SDW}} = {\cal O}_{f{\rm DW}}.
\label{eq:dualII_op}
\end{align}
Density waves of charges are transformed into spin currents while
density waves of spins into charge currents. This transformation
thus exchanges density and current as well as spin and charge
degrees of freedom. Similarly to $U_{I}(\theta)$, the unitary
operator $U_{II} (\theta)$ gives a continuous transformation
between dual operators such as
\begin{equation}
U_{II} (\theta){\cal O}_{\rm CDW}U_{II} (\theta)^{-1}= {\cal
O}_{\rm CDW}\cos \theta + {\cal O}_{d{\rm SDW}}\sin \theta.
\end{equation}

We note that under the transformation (\ref{eq:dual2}) the
$d$-wave and $s$-wave pairing operators, ${\cal O}_{d{\rm SC}}$
and ${\cal O}_{s{\rm SC}}$, are invariant except for phase
factors, and hence the $d$- and $s$-Mott states are also
invariant. On the other hand, the triplet parity-odd paring
operator ${\cal O}_{\rm t,o}$ is converted to the singlet
parity-odd one ${\cal O}_{\rm s,o}$,
\begin{equation}
\bar{\cal O}_{\rm t,o}={\cal O}_{\rm s,o}. \label{eq:dualII_op2}
\end{equation}

\section{Duality relations in the electronic ladder}
Here, we apply the duality transformations to the Hamiltonian
(\ref{eq:Ham}). It can be shown that these transformations map the
model onto the same Hubbard ladder with different coupling
parameters. It is convenient to rewrite the Hamiltonian in terms
of bonding and antibonding operators in the forms
\begin{align}
H_{0} &= - \sum_{j,\sigma} \Bigl[t_\parallel \sum_{\lambda=\pm}
(d_{j,\lambda,\sigma}^\dagger d_{j+1,\lambda,\sigma} +
\mbox{H.c.})
\nonumber\\
 & +t_\perp (d_{j,+,\sigma}^\dagger
d_{j,+,\sigma} - d_{j,-,\sigma}^\dagger d_{j,-,\sigma})\Bigr],
\end{align}
\begin{align}
\lefteqn{H_{\rm rung} = \sum_j \Bigl[ A
(d_{j,+,\uparrow}^\dagger d_{j,-,\uparrow}
d_{j,+,\downarrow}^\dagger d_{j,-,\downarrow} + {\rm H.c.})}
\nonumber\\
&+ B
(d_{j,+,\uparrow}^\dagger d_{j,-,\uparrow}
d_{j,-,\downarrow}^\dagger d_{j,+,\downarrow} + {\rm H.c.})
+ C\sum_{\sigma} n_{j,+,\sigma}^{(d)} n_{j,-,\sigma}^{(d)} \nonumber\\
&+ D
\sum_{\lambda=\pm} n_{j,\lambda,\uparrow}^{(d)}
n_{j,\lambda,\downarrow}^{(d)}
+ E
\sum_{\lambda=\pm} n_{j,\lambda,\uparrow}^{(d)}
n_{j,-\lambda,\downarrow}^{(d)} \Bigr],
\label{eq:Ham_trans}
\end{align}
where $n^{(d)}_{j,\lambda,\sigma}=d_{j,\lambda,\sigma}^\dagger
d_{j,\lambda,\sigma}$, and $\lambda =+$ ($-$) represents the
bonding (antibonding) orbital. The coupling constants in eq.\
(\ref{eq:Ham_trans}) are given by
\begin{align}
A &= (U - V_\perp + t_{\rm pair})/2 + (2J_\perp +
J^z_\perp)/8, \nonumber\\
B &= (U - V_\perp - t_{\rm pair})/2 -(2J_\perp -
J^z_\perp)/8, \nonumber\\
C &= V_\perp+J^z_\perp/4, \nonumber\\
D &= (U + V_\perp + t_{\rm pair})/2-(2J_\perp +
J^z_\perp)/8, \nonumber\\
E &= (U + V_\perp - t_{\rm pair})/2+(2J_\perp -
J^z_\perp)/8.\nonumber
\end{align}

Under the transformations (\ref{eq:dual1}) and (\ref{eq:dual2}),
the total charge density, total magnetization, and kinetic energy
term $H_0$ are invariant. Hence the chemical potential $\mu$,
magnetic field, and the parameters $t_\parallel$ and $t_\perp$ are
unchanged through the mapping.
On the other hand, the intra-rung coupling terms in $H_{\rm rung}$
are mixed up by the transformations. The parameter mappings are
given in the following.

\subsection{Duality relation I}
It is easy to see in eq.\ (\ref{eq:Ham_trans}) that the
transformation (\ref{eq:dual1}) changes only the sign of the
$A$-term, but keeps the rest of terms invariant. Hence, it gives a
one-to-one parameter mapping
\begin{equation}
(A,B,C,D,E) \rightarrow (-A,B,C,D,E).
\end{equation}
This leads to an exact duality relation in the parameter space of
the system. The model with a parameter $A$ is dual to the model
with $-A$ and self-dual in the space $A=0$, i.e.,
\begin{equation}
U - V_\perp + t_{\rm pair}+ (2J_\perp+J^z_\perp)/4 = 0.
\label{eq:self-dual}
\end{equation}
The coupling parameters are mapped as follows:
\begin{align}
\tilde{U} &= ( U + V_\perp - t_{\rm pair})/2 - (2 J_\perp +
J_\perp^z)/8,
\nonumber \\
\tilde{V}_\perp &= ( U + 3 V_\perp + t_{\rm pair} )/4 + (2J_\perp
+ J_\perp^z)/16,
\nonumber \\
\tilde{t}_{\rm pair} &= ( - U + V_\perp + t_{\rm pair} )/2 -
(2J_\perp + J_\perp^z)/8,
\label{eq:dual_paramI} \\
\tilde{J}_\perp &= - U + V_\perp - t_{\rm pair} + (2J_\perp -
J_\perp^z)/4 ,
\nonumber \\
\tilde{J}_\perp^z &= - U + V_\perp - t_{\rm pair} - (2 J_\perp - 3
J_\perp^z)/4. \nonumber
\end{align}
Note that the spin isotropy ($J_\perp=J_\perp^z$) is conserved
through this parameter mapping, i.e., $\tilde{J}_\perp =
\tilde{J}_\perp^z$.

{}From the mapping (\ref{eq:dual_paramI}) and the duality relation
(\ref{eq:dualityI}), one can conclude that if a density-wave
order, e.g., CDW or SDW order, appears in a certain parameter
region, a dual current order, i.e., $d$DW or $d$SDW order,
respectively, exists in a corresponding dual parameter region.
Because of this duality relation, all phase boundaries must be
symmetric with respect to the self-dual space. Indeed, the
transition line between the CDW and $d$DW phases derived for
half-filling and in the weak- and strong-coupling
limits\cite{LinBF,FjaerestadM,TsuchiizuF} coincides with the
self-dual line (\ref{eq:self-dual}).
We stress that our exact result holds in general cases, regardless
of the coupling strength, filling, and system size.

\subsection{Duality relation II}
From eq.\ (\ref{eq:Ham_trans}), one can see that the
transformation (\ref{eq:dual2}) changes only the sign of the
$B$-term and hence gives a one-to-one parameter mapping
\begin{equation}
(A,B,C,D,E) \rightarrow (A,-B,C,D,E).
\end{equation}
Thus the present model has another duality: The model with a
parameter $B$ is dual to the model with $-B$ and self-dual in the
space $B=0$, i.e.,
\begin{equation}\label{eq:self-dual3}
U-V_{\perp}-t_{\rm pair}-(2J_{\perp}-J_{\perp}^z)/4=0.
\end{equation}
The transformed coupling parameters are given by
\begin{align}
\bar{U} &= ( U + V_{\perp} + t_{\rm pair} )/2 +
(2J_{\perp}-J_{\perp}^z)/8,
\nonumber \\
\bar{V}_\perp &= ( U + 3V_{\perp} - t_{\rm pair} )/4 -
(2J_{\perp}-J_{\perp}^z)/16,
\nonumber \\
\bar{t}_{\rm pair} &= ( U - V_{\perp} + t_{\rm pair} )/2 -
(2J_{\perp}-J_{\perp}^z)/8,
\label{eq:dual_paramII} \\
\bar{J}_\perp &= U - V_{\perp} - t_{\rm pair} +
(2J_{\perp}+J_{\perp}^z)/4,
\nonumber \\
\bar{J}_\perp^z &= - U + V_{\perp} + t_{\rm pair} +
(2J_{\perp}+3J_{\perp}^z)/4.\nonumber
\end{align}

This duality relation leads to the conclusion that, if CDW or
$d$DW order, for example, appears in a certain parameter region,
spin current ($d$SDW) or SDW order exists in a dual region,
respectively. Note that even if we start from a spin isotropic
model ($J_\perp=J_\perp^z$), the dual model is spin anisotropic
and hence the spin symmetry breaking associated with SDW or $d$SDW
order can occur in the dual model. The whole phase diagram must be
symmetric with respect to the self-dual space
(\ref{eq:self-dual3}) and the direct phase transitions between
dual phases, if ever, locate exactly on the self-dual space.

\subsection{Duality relation between spin and charge}
A combination of the duality transformations (\ref{eq:dual1}) and
(\ref{eq:dual2}) leads to the spin-charge duality transformation
given by
\begin{equation}
\begin{array}{lc}
  \hat{d}_{j,+,\uparrow} = d_{j,+,\uparrow},~~~~~~~
  & \hat{d}_{j,+,\downarrow} = -i d_{j,+,\downarrow}, \\
  \hat{d}_{j,-,\uparrow} = d_{j,-,\uparrow},
  & \hat{d}_{j,-,\downarrow} = i d_{j,-,\downarrow},\\
\end{array}
\label{eq:dual_comb}
\end{equation}
which directly exchanges charge and spin degrees of freedom.
For example, general density-wave operators
$\sum_{{\bf k},\sigma}f_A({\bf k}) c^\dagger_\sigma({\bf k})
c_\sigma({\bf k+Q})$
are transformed to spin-density-wave operators
$\sum_{{\bf k}} \sum_{\sigma,\sigma'=\uparrow,\downarrow} f_A({\bf
k}) c^\dagger_\sigma({\bf k}) \sigma^z_{\sigma\sigma'}
c_{\sigma'}({\bf k+Q})$.


The transformation (\ref{eq:dual_comb}) gives another duality
relation in the Hamiltonian, which is given by the combination of
eqs.\ (\ref{eq:dual_paramI}) and (\ref{eq:dual_paramII}).
The model is self-dual in the space satisfying both
$U - V_{\perp} + J_{\perp}^z/4 = 0$
and
$t_{\rm pair} + J_{\perp}/2 = 0$,
which is the intersection of the self-dual spaces
(\ref{eq:self-dual}) for duality I and (\ref{eq:self-dual3}) for
duality II.

\subsection{Quantum phase transition at self-dual
points}\label{sec:criticality}
\begin{figure}[b]
  \centering
  \includegraphics[width=65mm]{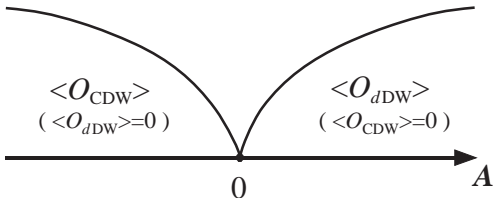}
  \caption{Orders of dual phases are destabilized at the self-dual points,
  i.e., $A=0$ for the duality I and $B=0$ for the duality II,
  because of the hidden U(1) symmetries. Hence the direct phase
  transitions between dual phases are of second order.
  }\label{fig:criticality}
\end{figure}
Here we briefly discuss the nature of quantum phase transitions at
self-dual points. In the self-dual spaces (\ref{eq:self-dual}) and
(\ref{eq:self-dual3}), the model Hamiltonian is invariant under
the continuous rotations given by the unitary operators
$U_I(\theta)$ and $U_{II}(\theta)$, respectively. [This can be
easily seen in eq.\ (\ref{eq:Ham_trans}) by setting $A=0$ or
$B=0$.] Hence the self-dual models have extra hidden U(1)
symmetries. Because of the U(1) symmetries in the self-dual space,
a rigorous theorem\cite{Momoi} concludes that the dual orders that
are continuously transformed to each other disappear on the
self-dual models in one dimension. Hence, the direct phase
transition between the dual phases, if it exists, must be of
second order. See Fig.\ \ref{fig:criticality}. An exception can
appear only if susceptibility of the generator for this rotation
is diverging at the transition (see ref.\ \citen{Momoi}); In this
case, the orders can survive even at the transition point in the
self-dual space, but there are gapless excitations associated with
the continuous U(1) symmetry breakdown. We note that the
discussion above does not exclude the presence of gapful phases in
the self-dual models. Self-dual phases, which can be either gapful
or gapless, may appear in a finite parameter region including the
self-dual space.

In the quantum phase transitions at self-dual points, the $A$- and
$B$-terms in the interaction (\ref{eq:Ham_trans}) serve as
symmetry-breaking perturbations for the U(1) symmetries associated
with $U_I(\theta)$ and $U_{II}(\theta)$, respectively. Actually,
the $A$- and $B$-terms can be expressed with the difference of
dual operators in the forms
\begin{align}
\lefteqn{A (d_{j,+,\uparrow}^\dagger d_{j,-,\uparrow}
d_{j,+,\downarrow}^\dagger d_{j,-,\downarrow} + {\rm H.c.})} \nonumber\\
&=A[({\cal O}_{\rm CDW})^2-({\cal O}_{d{\rm DW}})^2]
=A[({\cal O}_{d{\rm SDW}})^2-({\cal O}_{\rm SDW})^2] \nonumber\\
&=A[|{\cal O}_{s{\rm SC}}|^2-|{\cal O}_{d{\rm SC}}|^2],\\
\lefteqn{ B (d_{j,+,\uparrow}^\dagger d_{j,-,\uparrow}
d_{j,-,\downarrow}^\dagger d_{j,+,\downarrow} + {\rm H.c.}) }\nonumber\\
&=B[({\cal O}_{\rm CDW})^2-({\cal O}_{d{\rm SDW}})^2]
=B[({\cal O}_{d{\rm DW}})^2-({\cal O}_{\rm SDW})^2] \nonumber\\
&=B[|{\cal O}_{s,o}|^2-|{\cal O}_{t,o}|^2],
\end{align}
where $|{\cal O}|^2=({\rm Re}{\cal O})^2+({\rm Im}{\cal O})^2$.
When a symmetry-breaking perturbation is relevant, it induces a
order and a gap determining the criticality of the phase
transition (Fig.\ \ref{fig:criticality}). For example, the
transition between CDW and $d$DW phases was studied using the
Bosonization method in the weak-coupling limit and shown to be
characterized by $c=1$ Gaussian
criticality\cite{LinBF,TsuchiizuF}.

\section{The SO(5) symmetric ladder}
The SO(5) symmetric ladder presented by Scalapino {\it et
al.}\cite{ScalapinoZH} has only intra-rung couplings and belongs
to the Hamiltonian (\ref{eq:Ham}). They showed that the
half-filled Hubbard ladder ($\mu=\frac{1}{2}U+V$) has the SO(5)
symmetry if the parameters satisfy
\begin{equation}
J_\perp = J_\perp^z = 4 ( U + V_\perp ). \label{eq:SO5}
\end{equation}
No condition is imposed on the values of $t_{\perp}$,
$t_{\parallel}$, and $t_{\rm pair}$, since the hopping terms in
$H_0$ and the pair-hopping term ($t_{\rm pair}$) are SO(5)
symmetric\cite{ScalapinoZH,FrahmS}. In this section, we show a
duality relation in the SO(5) symmetric ladder.

Scalapino {\it et al.}\cite{ScalapinoZH} showed that the
five-dimensional SO(5) superspin vector $n^a$ ($a=1,\dots,5$) is
related to the AFM and $d$SC operators by
\begin{align}
n^1 &= \sqrt{2}{\rm Re}{{\cal O}_{d{\rm SC}}},~~~~~~~
n^{(2,3,4)} = S_1^{(x,y,z)} - S_2^{(x,y,z)},
\nonumber\\
n^5 &= \sqrt{2}{\rm Im}{{\cal O}_{d{\rm SC}}},\label{eq:SO(5)spin}
\end{align}
where we have omitted rung indices $j$. In terms of the rung
charge $Q=\frac{1}{2}\sum_\sigma(n_{1\sigma} + n_{2\sigma} -1)$,
the rung spin $S^\alpha=S_1^\alpha+S_2^\alpha$, and the
$\pi_\alpha$ operators
$\pi_\alpha^\dagger = - \frac{1}{2} \sum_{\sigma,\sigma'}
c^\dagger_{1\sigma}(\sigma_\alpha \sigma_y)_{\sigma\sigma'}
c^\dagger_{2\sigma'}$,
the ten-dimensional SO(5) symmetry generators $L^{ab}$
($a,b=1,\dots,5$) are expressed with
\begin{equation}
\left[%
\begin{array}{ccccc}
  0 &  &  &  &  \\
  \pi_x^\dagger+\pi_x & 0 &  &  &  \\
  \pi_y^\dagger+\pi_y & -S^z & 0 &  &  \\
  \pi_z^\dagger+\pi_z & S^y & -S^x & 0 &  \\
  Q & \frac{1}{i}(\pi_x^\dagger-\pi_x) & \frac{1}{i}(\pi_y^\dagger-\pi_y)
  & \frac{1}{i}(\pi_z^\dagger-\pi_z) & 0 \\
\end{array}%
\right],
\label{eq:SO(5)generator}
\end{equation}
where the elements are antisymmetric. The SO(5) scalar operator
$\rho$ is expressed by the charge density operator with
\begin{equation}
\rho=
{\cal O}_{\rm CDW}+1.
\end{equation}

Applying the duality transformation I, we find that the
transformed superspin vector $\tilde{n}^a$ is related to the spin
current ($d$SDW) and $s$SC operators by
\begin{equation}
\begin{split}
\tilde{n}^1 &= \sqrt{2}{\rm Re}{{\cal O}_{s{\rm SC}}}, ~~~~~~~~~
\tilde{n}^5 = \sqrt{2}{\rm Im}{{\cal O}_{s{\rm SC}}},\\
\tilde{n}^{(2,3,4)} &= -\frac{i}{2}\sum_{\sigma,\sigma'}
\sigma_{\sigma \sigma'}^{(x,y,z)} (c_{1,\sigma}^\dagger
c_{2,\sigma'} - {\rm H.c.})=j_s^{(x,y,z)},
\end{split}
\end{equation}
where the three elements $j_s^\alpha$ ($\alpha=x,y,z$) are
Cartesian components of the spin current ($d$SDW) along each rung
and $j_s^z={\cal O}_{d{\rm SDW}}$. Moreover, using the relations
$\tilde{\bm S}={\bm S}$, $\tilde{\bm \pi}=i {\bm \pi}$, and
$\tilde{Q}=Q$, one can find that the transformed symmetry
generators $\tilde{L}^{ab}$ are given by just a permutation of
original generators $L^{ab}$,
\begin{equation}
\begin{split}
\tilde{L}^{(2,3,4)1}=L^{5(2,3,4)},~~~~~~~
\tilde{L}^{5(2,3,4)}=-L^{(2,3,4)1},
\end{split}
\end{equation}
and the rest of $L^{ab}$ ($a \ge b$) are unchanged.
Thus, the set of SO(5) generators are invariant under the duality
transformation I, except for the permutation of elements. From
these results, we can conclude that, besides the superspin vector
(\ref{eq:SO(5)spin}), the same SO(5) symmetry with the generators
(\ref{eq:SO(5)generator}) unifies the $d$SDW and $s$SC in another
grand order parameter field
\begin{equation}
(\sqrt{2}{\rm Im}{\cal O}_{s{\rm SC}},j_s^x,j_s^y,j_s^z,
-\sqrt{2}{\rm Re}{\cal O}_{s{\rm SC}}). \label{eq:SO5vec2}
\end{equation}
The transformed SO(5) scalar operator $\tilde{\rho}$ is given by
the $d$DW operator with
\begin{equation}
\tilde{\rho}=
-{\cal O}_{d{\rm DW}}+1.
\end{equation}
One may be surprised that the $d$DW operator is an SO(5) scalar,
but one can easily express the $d$DW (and CDW) operator in terms
of SO(5) spinors in an SO(5) symmetric form.

Since the set of SO(5) generators are invariant under the duality
transformation I, the SO(5) symmetric Hubbard ladder is mapped
onto the SO(5) symmetric model satisfying the condition
(\ref{eq:SO5}). One can easily check that the condition
(\ref{eq:SO5}) is conserved under the parameter mapping
(\ref{eq:dual_paramI}). However, the parameter point in
($J_\perp,U,t_{pair}$) for the SO(5) symmetric model is converted
to the dual point with
\begin{align}
\tilde{U} &= - U - V_\perp - t_{\rm pair}/2,~~~~~~~
\tilde{J}_\perp = 2 V_\perp - t_{\rm pair},
\nonumber \\
\tilde{t}_{\rm pair} &= -2 U -V_\perp + t_{\rm pair}/2,
\end{align}
and $\tilde{V}_\perp$ is given by
$\tilde{V}_\perp=\tilde{J}_\perp/4-\tilde{U}$.

To demonstrate the duality relation in the SO(5) symmetric model,
we write down the Hamiltonian in terms of dual and self-dual
operators. The intra-rung part (\ref{eq:Hrung}) can be cast into
the form
\begin{equation}
\begin{split}
\lefteqn{H_{\rm rung} -\mu_0\sum_{j,l,\sigma} n_{j,l,\sigma}
=\sum_j\left[\left(\frac{J_\perp}{4}
-\frac{t_{\rm pair}}{2}\right)\sum_{a<b}(L^{ab}_j)^2\right.} \\
&+\left.\left(\frac{J_\perp}{2}+2U-t_{\rm pair}\right)(\rho_j-1)^2
-2t_{\rm pair}(\tilde{\rho}_j-1)^2\right],
\end{split}\label{eq:Hdual1}
\end{equation}
where $\mu_0=\frac{1}{2}U+V$. The Casimir operator
$C=\sum_{a<b}(L^{ab})^2$ is invariant under the duality
transformation, and the kinetic energy term as well. It is also
instructive to rewrite the Hamiltonian in terms of the dual
superspin vectors in the form
\begin{align}
\lefteqn{H_{\rm rung}-\mu_0\sum_{j,l,\sigma}
n_{j,l,\sigma}}\nonumber\\
&= \sum_j \left[
\left(\frac{J_\perp}{20}-\frac{4U}{5}+\frac{7t_{\rm
pair}}{10}\right) \sum_{a<b}(L^{ab}_j)^2\right.
\label{eq:Hdual2}\\
&-\left.\frac{1}{10}\left(J_\perp+4U-2t_{\rm
pair}\right)\sum_a(n^a_j)^2 + \frac{2}{5}t_{\rm
pair}\sum_a(\tilde{n}_j^a)^2\right], \nonumber
\end{align}
where we have used the Fierz identity\cite{ScalapinoZH}
\begin{equation}
5 (\rho-1)^2 = 5 - ({\bm n})^2 - 2 \sum_{a<b} (L^{ab})^2.
\end{equation}
The duality relation in the Hamiltonian is transparent in the
forms (\ref{eq:Hdual1}) and (\ref{eq:Hdual2}). The SO(5) symmetric
Hamiltonian is self dual if two coefficients of dual operators are
equal, i.e.,
\begin{equation}
J_\perp+4U+2t_{\rm pair}=0. \label{eq:SO(5)selfdual}
\end{equation}
The self-dual SO(5) model has a superspin SO(5) $\times$ duality
U(1) symmetry, in total. The two-dimensional self-dual plane
divides the three-dimensional SO(5) symmetric parameter space into
two subspaces. In one subspace, the AFM and $d$SC correlations are
dominant, showing a crossover upon doping as was discussed in
refs.\ \citen{SCZhang,RabelloKDZ,Henley,ScalapinoZH}. In the other
subspace, the $d$SDW and $s$SC correlations come out to be
relevant orders and hence symmetry breaking perturbations enhance
$d$SDW or $s$SC order in the same way as the case of the AFM and
$d$SC orders.

For $t_{\rm pair} = 0$ and $U < 0$, this self-dual plane
(\ref{eq:SO(5)selfdual}) coincides with a phase boundary line
found in the previous
studies.\cite{MarstonFS,ScalapinoZH,LinBF,FrahmS,FjaerestadM}
For example, the transition line between the CDW and $d$DW phases
given in ref.\ \citen{MarstonFS} is identical with the self-dual
line. For $t_{\rm pair} = 0$ and $U > 0$, on the other hand, the
self-dual space locates in the C2S2 critical phase. This implies
that the critical phase is self-dual and a crossover between dual
states characterized by different dominant correlations occurs on
this self-dual line.

\section{Duality relations in weak coupling}
Here we reconsider the duality relations in the weak-coupling
limit. In this limit, the duality relation I corresponds to an
already known relation in Bosonization
studies\cite{LinBF,FjaerestadM,TsuchiizuF}. We also find another
new duality relation which appears only in the low-energy
effective theory.
\subsection{Bosonization framework}
We apply the Abelian bosonization method.\cite{Tsvelik,BalentsF}
In a continuum limit, we expand the electron-field operators as
\begin{equation}
 \psi_{\lambda \sigma}(x)
 = e^{ ik_{F\lambda}x} \psi_{R \lambda \sigma}(x)
 + e^{-ik_{F\lambda}x} \psi_{L \lambda \sigma}(x)
\end{equation}
for $\lambda = \pm$ and $\sigma=\uparrow,\downarrow$, where
$\psi_{R \lambda \sigma}$ ($\psi_{L \lambda \sigma}$) represents
the chiral fields for right- (left-) moving electrons in the
bonding ($\lambda=+$) and antibonding ($\lambda=-$) bands and
$k_{F\lambda}$ are the Fermi wave vectors.
As in the usual way, we introduce boson fields $\varphi_{p \lambda
\sigma}$ for the chiral fields as
\begin{equation}
 \psi_{p \lambda \sigma}(x) =
 \frac{\eta_{\lambda \sigma}}{\sqrt{2\pi a_0}}
 \exp [ ip\varphi_{p \lambda\sigma}(x) ],
\end{equation}
for $p = R/L = +/-$, where $\eta_{\lambda\sigma}$ denote the Klein
factors and $a_0$ the lattice constant. Then, we define a new set
of boson fields:
\begin{equation}
\phi_{\nu r}=\phi^R_{\nu r} + \phi^L_{\nu r},~~~~~~~ \theta_{\nu
r}=\phi^R_{\nu r} - \phi^L_{\nu r}, \label{eq:phitheta}
\end{equation}
with
\begin{subequations}
\begin{align}
 \phi^p_{c r} &= \frac{1}{4}
 \{\varphi_{p + \uparrow} + \varphi_{p + \downarrow}
 + r (\varphi_{p - \uparrow} + \varphi_{p - \downarrow})\},\\
 \phi^p_{s r} &= \frac{1}{4}
 \{\varphi_{p + \uparrow} - \varphi_{p + \downarrow}
 + r (\varphi_{p - \uparrow} - \varphi_{p - \downarrow})\}
\end{align}
\end{subequations}
for $\nu = c, s$ and $r=\pm$. The $\phi$ and $\theta$ fields
(\ref{eq:phitheta}) satisfy the commutation relations $[\phi_{\nu
r}(x), \theta_{\nu' r'}(x')] = -i\pi\Theta(x'-x)\delta_{\nu
\nu'}\delta_{r r'}$ [$\Theta(x)$ is the Heaviside step function]
and $[\phi_{\nu r}(x), \phi_{\nu' r'}(x')] = [\theta_{\nu r}(x),
\theta_{\nu' r'}(x')] = 0$.

Using these boson fields, one finds that the duality
transformation I is expressed as the translation of the
$\theta_{cr}$ fields by $r\pi/2$ while the duality transformation
II is the translation of the $\theta_{c+}$ and $\theta_{s-}$
fields by $\pi/2$. It was already realized that the former
transformation gives a duality relation in the Bosonization
method\cite{LinBF,FjaerestadM,TsuchiizuF}.

\subsection{Duality relation III}
Here we restrict our discussion to the half-filled system.
Applying gauge transformations to right-moving and left-moving
chiral fermion fields independently, we can form another duality
transformation
\begin{equation}
\check{\psi}_{R\lambda\sigma} =
 e^{i\pi/4} \psi_{R\lambda\sigma},~~~~~~
\check{\psi}_{L\lambda\sigma} =
 e^{-i\pi/4} \psi_{L\lambda\sigma},
\label{eq:dual3}
\end{equation}
which gives the following relations between order parameters
\begin{equation}
\begin{split}
\check{O}_{\rm CDW}(\pi) &\sim - {O}_{p{\rm DW}}(\pi), ~~~~~
\check{O}_{d{\rm DW}}(\pi) \sim {O}_{f{\rm DW}}(\pi),\\
\check{O}_{\rm SDW}(\pi) &\sim - {O}_{p{\rm SDW}}(\pi), ~~~~~
\check{O}_{d{\rm SDW}}(\pi) \sim {O}_{f{\rm SDW}}(\pi),
\end{split}
\label{eq:dualityIII}
\end{equation}
where we have omitted $k_{F\lambda}$-dependent prefactors. We
thus find that the transformation (\ref{eq:dual3}) exchanges
rung-centered orders ($s$- and $d$-wave orders) and
plaquette-centered orders ($p$- and $f$-wave orders). See Fig.\
\ref{fig:duality}. The low-energy effective Hamiltonian can be
decoupled into self-dual parts and symmetry breaking
perturbations, in which the duality relation is visible. The
duality relation III thereby gives a parameter mapping as similar
to the duality relations I and II. In terms of the bosonic fields,
this transformation is given by the translation of the $\phi_{c+}$
field by $\pi/2$ and
one can see that the duality relation (\ref{eq:dualityIII}) was
transparent in the results of the bosonization
studies\cite{FjaerestadM,TsuchiizuF}. We note that this
transformation can be written in such a compact form only in the
low-energy effective theory.

\subsection{Ising duality}
Finally we mention the Ising duality transformation\cite{LinBF},
which interchanges the $\phi_{s-}$ and $\theta_{s-}$ fields,
\begin{equation}
\theta_{s -} \to \phi_{s -}, ~~~~~~~
\phi_{s -} \to \theta_{s -},
\end{equation}
and leaves the rest of the $\phi$ and $\theta$ fields unchanged.
This transformation converts a $d$DW state into a $d$-Mott state
and a CDW state into an $s$-Mott state. Thus, this is a duality
between ordered and disordered phases and qualitatively different
from the duality relations I-III.

\section{Strong coupling approach at half-filling}
In this section, we discuss the stability of ordered states at
half-filling using strong-coupling expansions as was discussed in
ref.\ \citen{TsuchiizuF}. Using the duality transformations I and
II, we can easily find various transition lines between ordered
(density-wave) and disordered (Mott-insulating) phases.

Let us discuss CDW, $d$DW, SDW, and $d$SDW orders. For individual
phases, an effective Hamiltonian can be derived on doubly
degenerate rung basis $|+\rangle$ and $|-\rangle$ in the form
\begin{equation}
H_{\rm eff}=\sum_j (K\tau_j^z \tau_{j+1}^z + h \tau_j^x ),
\end{equation}
where $\tau_j^\alpha$ ($\alpha=x$, $y$, $z$) denotes the Pauli
matrices, and $\tau^z_j |\pm\rangle=\pm|\pm\rangle$. This
Hamiltonian suggests that doubly degenerate ordered ground states
appear in $K>|h|$ and disordered states in $K<|h|$. These
transitions are characterized by the Ising criticality ($c=1/2$).

In the following derivation, we use convenient basis,
\begin{equation}
\begin{split}
|1\rangle &\equiv
c_{j,1,\uparrow}^\dagger c_{j,2,\downarrow}^\dagger
|0\rangle,~~~~~~ |2\rangle \equiv
c_{j,1,\downarrow}^\dagger c_{j,2,\uparrow}^\dagger |0\rangle,
\\
|3\rangle &\equiv
c_{j,1,\uparrow}^\dagger c_{j,1,\downarrow}^\dagger
|0\rangle,~~~~~~ |4\rangle \equiv
c_{j,2,\uparrow}^\dagger c_{j,2,\downarrow}^\dagger |0\rangle.
\end{split}
\end{equation}

\subsection{CDW--$s$-Mott}
The perturbation expansion for a CDW ordered state can be
performed in the strong coupling limit $2V_\perp - U\rightarrow
\infty$ and $-U-J_\perp/2-J_\perp^z/4+V_\perp \rightarrow \infty$,
where two degenerate ground-state basis per rung are
$|+\rangle = |3\rangle$ and $|-\rangle = |4\rangle$.
By treating hopping terms ($t_\parallel,t_\perp,t_{\rm pair}$) as
the perturbation, the couplings of the effective Hamiltonian are
derived as
\begin{subequations}
\begin{align}
K &= 2{t_\parallel}^2/(2 V_\perp - U),\\
h &= t_{\rm pair}
-2{t_\perp}^2/(-U+V_\perp-J_\perp/2-J_\perp^z/4).
\end{align}
\end{subequations}
A CDW ordered state appears for $K>|h|$ and an $s$-Mott state for
$K< |h|$.

\subsection{$d$DW--$d$-Mott}
Using the duality transformations, we can derive the effective
Hamiltonian for other ordered states. For the $d$DW phase, the
unperturbed ground-state basis are
$|\pm\rangle = \frac{1}{2}\{(|1\rangle-|2\rangle)\mp i( |3\rangle
- |4\rangle)\}$
in the strong coupling limit $V_\perp+t_{\rm pair} +
J_\perp/2+J_\perp^z/4 \rightarrow \infty$ and $U-V_\perp+ 3t_{\rm
pair} + J_\perp/2+J_\perp^z/4 \rightarrow \infty$. The couplings
of the effective Hamiltonian are given by
\begin{subequations}
\begin{align}
K &= 8{t_\parallel}^2/(4V_\perp+ 4t_{\rm pair} +
2J_\perp + J_\perp^z),\\
h &= (-U+V_\perp+t_{\rm pair})/2 - (2J_\perp+J_\perp^z)/8
\nonumber\\
&- 4{t_\perp}^2/(U-V_\perp+ 3t_{\rm pair} +
J_\perp/2+J_\perp^z/4).
\end{align}
\end{subequations} One can say that a $d$DW ordered state
appears for $K>|h|$ and a $d$-Mott state for $K< |h|$.

\subsection{SDW--$d$-Mott}
For the SDW phase,
$|+\rangle = |1\rangle$ and $|-\rangle = -|2\rangle$
in the strong coupling limit $2U+J_\perp^z \rightarrow\infty$ and
$U-V_\perp+t_{\rm pair}+J_\perp^z/4\rightarrow\infty$. The
couplings of the effective Hamiltonian are derived as
\begin{subequations}
\begin{align}
K &= 4{t_\parallel}^2/(J_\perp^z+2U),\\
h &= - J_\perp/2 - 2{t_\perp}^2/(U-V_\perp+t_{\rm
pair}+J_\perp^z/4).
\end{align}
\end{subequations} One can say that an SDW ordered state
appears for $K>|h|$ and a $d$-Mott state for $K<|h|$.

\subsection{$d$SDW--$s$-Mott}
For the $d$SDW phase,
$|\pm \rangle = \frac{1}{2}\{(|3\rangle+|4\rangle) \pm i(
|1\rangle + |2\rangle)\}$
in the strong coupling limit $V_\perp-t_{\rm pair} -
J_\perp/2+J_\perp^z/4\rightarrow\infty$ and $- U + V_\perp -
t_{\rm pair} - 3J_\perp/2 - J_\perp^z/4
\rightarrow\infty$. The couplings of the effective Hamiltonian are
derived as
\begin{subequations}
\begin{align}
K &= 8{t_\parallel}^2/(4V_\perp-4t_{\rm pair} -
2J_\perp + J_\perp^z),\\
h &= (U-V_\perp+t_{\rm pair})/2 - (2J_\perp-J_\perp^z)/8
\nonumber\\
& - 4{t_\perp}^2/(- U + V_\perp - t_{\rm pair} - 3J_\perp/2 -
J_\perp^z/4).
\end{align}
\end{subequations} One can say that a $d$SDW (spin current)
ordered state appears for $K>|h|$ and an $s$-Mott state for $K<
|h|$.

\section{Discussions}
In summary, we have established duality relations in correlated
electron systems on the two-leg ladder. Our arguments clarify
mutual relations between conventional and various unconventional
density-wave orders, as shown in Fig.\ \ref{fig:duality}. The
present duality arguments reveal that the stability of
unconventional density-wave orders such as staggered flux ($d$DW)
and circulating spin current ($d$SDW) is equal to that of
conventional density-wave orders in the dual parameter spaces.
Recently, large-scale numerical analyses\cite{Schollwoeck}
reported the appearance of incommensurate $d$DW (quasi-)long range
order upon doping. Applying the duality relations to this result,
we can immediately conclude that new incommensurate CDW, SDW, and
$d$SDW phases stably exist in the doped generalized Hubbard ladder
in the dual parameter spaces.
These duality arguments can be easily generalized to the Hubbard
ladder including inter-rung interactions.

Next, we found a duality structure in the SO(5) symmetric Hubbard
ladder system. The SO(5) symmetry, which was proposed to unify AFM
and $d$SC, also unifies $d$SDW and $s$SC. This gives a new route
to $s$-wave superconductivity in strongly correlated electron
systems. If the coupling parameters of a half-filled system is
close to the SO(5) symmetric region and the circulating spin
current correlation is dominant in the ground state, hole doping
can be a symmetry breaking perturbation which enhances the
$s$-wave superconductivity. Thus this system would show a
crossover from a $d$SDW dominant state to a $s$SC one upon doping.

Finally, we give a remark on the spin-chirality duality which was
previously introduced for the spin
ladder\cite{HikiharaMH,MomoiHNH}. In the spin-chirality duality
transformation, antiferromagnetic spin and vector chirality
degrees of freedom are converted to each other. One can see that
this spin-chirality duality has an analogy with the duality I for
electron systems: in the duality I, SDW (antiferromagnetic spin
order) is related to spin current, which is expected to have a
spin vector chiral order. Actually, Lecheminant and
Totsuka\cite{LecheminantT} showed in a Majorana fermion
representation that the spin-chirality duality transformation can
be written as a gauge transformation for the fermions, which is
similar to the transformation given in the present paper.


\acknowledgments The authors would like to thank M.\ Tsuchiizu,
H.\ Tsunetsugu, A.\ Furusaki, Y.\ Ohashi, and K.\ Totsuka for
stimulating discussions. They are also grateful to M.\ Nakamura
for useful comments on the unitary representation of the duality
transformations.

\end{document}